\documentclass[12pt]{article}
\usepackage[utf8]{inputenc} 

\usepackage{cite} 
\usepackage{url}  
\usepackage{ifthen}  
\usepackage{multicol}   
\usepackage[hidelinks]{hyperref}
\usepackage{amsmath}
 \usepackage{amsfonts,graphics,pdfpages}

\begin{document}

\title{Differential meta-analysis of RNA-seq data from multiple studies}

\author{Andrea Rau$^{1,2}$%
	\and
	Guillemette Marot$^{3,4}$%
	\and
	Florence Jaffr{\'e}zic$^{1,2}$%
}

\date{June 13, 2013}

\maketitle

{\small{
\begin{center}
\begin{enumerate}
\item INRA, UMR1313 G{\'e}n{\'e}tique animale et biologie int{\'e}grative, 78352 Jouy-enJosas, France\\
\item AgroParisTech, UMR1313 G{\'e}n{\'e}tique animale et biologie int{\'e}grative, 75231 Paris 05, France\\
\item Universit\'e Lille Nord de France, UDSL, EA2694 Biostatistics\\
\item Inria Lille Nord Europe, MODAL
\end{enumerate}
\end{center}
}

{\section*{ABSTRACT}}
{\bf Background:} High-throughput sequencing is now regularly used for studies of the transcriptome (RNA-seq), particularly for comparisons among experimental conditions. For the time being, a limited number of biological replicates are typically considered in such experiments, leading to low detection power for differential expression. As their cost continues to decrease, it is likely that additional follow-up studies will be conducted to re-address the same biological question.
\\
{\bf Results:} We demonstrate how $p$-value combination techniques previously used for microarray meta-analyses can be used for the differential analysis of RNA-seq data from multiple related studies. These techniques are compared to a negative binomial generalized linear model (GLM) including a fixed study effect on simulated data and real data on human melanoma cell lines. The GLM with fixed study effect performed well for low inter-study variation and small numbers of studies, but was outperformed by the meta-analysis methods for moderate to large inter-study variability and larger numbers of studies. 
\\
{\bf Conclusions:} The $p$-value combination techniques illustrated here are a valuable tool to perform differential meta-analyses of RNA-seq data by appropriately accounting for biological and technical variability within studies as well as additional study-specific effects. An R package \href{https://r-forge.r-project.org/R/?group_id=1504}{\texttt{metaRNASeq}} is available on the R Forge.
\\
{\bf Keywords}: meta-analysis, RNA-seq, differential expression, $p$-value combination

\newpage

\section{Background}

Studies of gene expression have come to rely increasingly on the use of high-throughput sequencing (HTS) techniques to directly sequence libraries of reads (i.e., nucleotide sequences) arising from the transcriptome (RNA-seq), yielding counts of the number of reads arising from each gene. Although the costs of HTS experiments continue to decrease, for the time being RNA-seq experiments are typically performed on very few biological replicates, and therefore analyses to detect differential expression between two experimental conditions tend to lack detection power. However, as costs continue to decrease, it is likely that additional follow-up experiments will be conducted to re-address some biological questions, suggesting a future need for methods able to jointly analyze data from multiple studies. In particular, such methods must be able to appropriately account for the biological and technical variability among samples within a given study as well as for the additional variability due to study-specific effects. Such inter-study variability may arise due to technical differences among studies (e.g., sample preparation, library protocols, batch effects) as well as additional biological variability. 

In recent years, several methods have been proposed to analyze microarray data arising from multiple independent but related studies; these meta-analysis techniques have the advantage of increasing the available sample size by integrating related datasets, subsequently increasing the power to detect differential expression. Such meta-analyses include, for example, methods to combine $p$-values \cite{Marot2009}, estimate and combine effect sizes \cite{Choi2003}, and rank genes within each study \cite{Breitling2004}; see \cite{Hu2006} and \cite{Hong2008} for a review and comparison of such methods. In particular, \cite{Marot2009} showed that the inverse normal $p$-value combination technique outperformed effect size combination methods or moderated $t$-tests \cite{limma} obtained from a linear model with a fixed study effect on several criteria, including sensitivity, area under the Receiver Operating Characteristic (ROC) curve, and gene ranking. 

In many cases the meta-analysis techniques previously used for microarray data are not directly applicable for RNA-seq data. In particular, differential analyses of microarray data, whether for one or multiple studies, typically make use of a standard or moderated $t$-test \cite{limma,SMVar}, as such data are continuous and may be roughly approximated by a Gaussian distribution after log-transformation. On the other hand, the growing body of work concerning the differential analysis of RNA-seq data has primarily focused on the use of overdispersed Poisson \cite{Auer2011} or negative binomial models \cite{DESeq, Robinson2010a} in order to account for their highly heterogeneous and discrete nature. Under these models, the calculation and interpretation of effect sizes is not straightforward. \cite{Kulinskaya2008} recently proposed an effect size combination method for Poisson-distributed data, based on an Anscombe transformation, but this method is not well-adapted to RNA-seq data due to the presence of over-dispersion among biological replicates as well as zero-inflation. To our knowledge, no other transformation has been proposed to obtain effect sizes for over-dispersed Poisson or negative binomial data.

In this paper, we consider several methods for the integrated analysis of RNA-seq data arising from multiple related studies, including two $p$-value combination methods as well as a model fitted over the full data with a fixed study effect. We first demonstrate how the inverse normal and Fisher $p$-value combination methods can be adapted to the differential meta-analysis of RNA-seq data. Then we compare these two methods to the results of independent per-study analyses and a negative binomial generalized linear model (GLM) with a fixed study effect as implemented in the \texttt{DESeq} Bioconductor package \cite{DESeq}. All methods are compared on real data from two related studies on human melanoma cell lines, as well as in an extensive set of simulations varying the inter-study variability, number of studies, and biological replicates per study.

Finally, we note that our focus is on RNA-seq data arising from two or more studies in which all experimental conditions under consideration are included in every study (with potentially different numbers of biological replicates); differential analyses among conditions that are not studied in the same experiment are typically limited, or even compromised, due to the confounding of condition and study effects.

\section{Methods}

Let $y_{gcrs}$ be the observed count for gene $g$ ($g = 1,\ldots,G$), condition $c$ ($c = 1,2$), biological replicate $r$ ($r = 1,\ldots R_{cs}$), and study $s$ ($s = 1,\ldots,S$). Note that the number of biological replicates $R_{cs}$ may vary between conditions and among studies. Let $\mu_{gcs}$ be the mean expression level for gene $g$ in condition $c$ and study $s$. For an integrated differential analysis of gene expression across all studies, two approaches can be envisaged: the combination of $p$-values from per-study differential analyses, and a global differential analysis. We illustrate both using the default methods and parameters of the \texttt{DESeq} (v1.10.1)  analysis pipeline \cite{DESeq}, although other popular methods, e.g., \texttt{edgeR} \cite{Robinson2010a}, could also be used.

\subsection{$P$-value combination from independent analyses}\label{sec:independent}

For the differential analysis of gene expression within a given study $s$, we assume that gene counts $y_{gcrs}$ follow a negative binomial distribution parameterized by its mean $\ell_{crs}\mu_{gcs}$ and dispersion $\alpha_{gs}$, where $\ell_{crs}$ is a normalization factor to correct for differences in library size. A comparison of different methods to estimate $\ell_{crs}$ may be found in \cite{Statomique2012}. We are interested in testing the per-gene null hypothesis 
\begin{align}
H_{0,gs}: \mu_{g1s} = \mu_{g2s} \hspace*{.25cm} \text{  vs  } \hspace*{.25cm} H_{1,gs}: \mu_{g1s} \ne \mu_{g2s}. \label{eqn:nullhyp}
\end{align}
After obtaining per-gene mean and dispersion parameter estimates, a parametric gamma regression is used to obtain fitted dispersion estimates by pooling information from genes with similar expression strengths. Raw per-gene $p$-values $p_{gs}$ are subsequently computed using a conditioned test analogous to Fisher's exact test. Additional details may be found in \cite{DESeq} and the \texttt{DESeq} package vignette.  Once these vectors of raw $p$-values have been obtained, we consider two possible approaches to combine them across studies: the inverse normal and the Fisher combination methods. We note that both of these approaches assume that under the null hypothesis, each vector of $p$-values is assumed to be uniformly distributed.

\subsubsection{Inverse normal method}

For each gene $g$, we define
\begin{equation} 
N_g = \sum_{s=1}^S w_s \Phi^{-1}(1-p_{gs}) \label{eqn:InvNorm}
\end{equation} 
where $p_{gs}$ corresponds to the raw $p$-value obtained for gene $g$ in a differential analysis for study $s$, $\Phi$ the cumulative distribution function of the standard normal distribution, and $w_s$ a set of weights \cite{Liptak1958}. We propose here to define the study-specific weights $w_s$ as in \cite{MarotMayer2009}:
\begin{equation} 
w_s=\sqrt{\frac{\sum_{c} R_{cs}}{\sum_{\ell}\sum_{c} R_{c\ell}}}, \nonumber
\end{equation} 
where $\sum_{c} R_{cs}$ is the total number of biological replicates in study $s$. This allows studies with large numbers of biological replicates to be attributed a larger weight than smaller studies. We note that other weights may also be defined by the user depending on the quality of the data in each study, if this information is available.

Under the null hypothesis, the test statistic $N_g$ in Equation~(\ref{eqn:InvNorm}) follows a ${\mathcal{N}}(0,1)$ distribution. A unilateral test on the right-hand tail of the distribution may then be performed, and classical procedures for the correction of multiple testing such as \cite{BH} may subsequently be applied to the obtained $p$-values to control the false discovery rate at a desired level $\alpha$.

\subsubsection{Fisher combination method}

For the Fisher combination method \cite{Fisher1932}, the test statistic for each gene $g$ may be defined as
\begin{equation} 
F_g = -2 \sum_{s=1}^S \ln (p_{gs}), \label{eqn:Fisher}
\end{equation}
where as before $p_{gs}$ corresponds to the raw $p$-value obtained for gene $g$ in a differential analysis for study $s$. Under the null hypothesis, the test statistic $F_g$ in Equation~(\ref{eqn:Fisher}) follows a $\chi^2$ distribution with $2S$ degrees of freedom. As with the 
inverse normal $p$-value combination method, classical procedures for the correction of multiple testing such as \cite{BH} may be applied to the obtained $p$-values to control the false discovery rate at a desired level $\alpha$.

\subsubsection{Additional considerations for $p$-value combination}\label{sec:considerations}

We note that the implementation of the previously described $p$-value combination techniques requires two additional considerations to be taken into account. 

First, a crucial underlying assumption for the statistics defined in Equations~(\ref{eqn:InvNorm}) and (\ref{eqn:Fisher}) is that $p$-values for all genes arising from the per-study differential analyses are uniformly distributed under the null hypothesis $H_{0,gs}$ defined in Equation~(\ref{eqn:nullhyp}). This assumption is, however, not always satisfied for RNA-seq data; in particular, a peak is often observed for $p$-values close to 1 due to the discretization of $p$-values for very low counts. To circumvent this first difficulty, we propose to filter the weakly expressed genes in each study using the \texttt{HTSFilter}  Bioconductor package \cite{Rau2012}, as described in the Supplementary Materials. We note that in so doing, it is possible for a gene to be filtered from one study and not from another. As will be seen in the following, this approach appears to effectively filter those genes contributing to a peak of large $p$-values, resulting in $p$-values that appear to be roughly uniformly distributed under the null hypothesis.

Second, for the two $p$-value combination methods described above, unlike microarray data, under- and over-expressed genes are analyzed together for RNA-seq data. As such, some care must be taken to identify genes exhibiting conflicting expression patterns (i.e., under-expression when comparing one condition to another in one study, and over-expression for the same comparison in another study). We suggest that genes exhibiting differential expression conflicts among studies be identified post hoc, and removed from the list of differentially expressed genes; this step to remove genes with conflicting differential expression from the final list of differentially expressed genes may be performed automatically within the associated R package \texttt{metaRNASeq}. 

\subsection{Global differential analysis}

For a global analysis of RNA-seq data arising from multiple studies, we assume that gene counts $y_{gcrs}$ follow a negative binomial distribution parameterized by the mean $\ell_{crs}\mu_{gcs}$ and dispersion $\alpha_{g}$, where $\ell_{crs}$ is the library size normalization factor. We are now interested in testing the global per-gene null hypothesis 
\begin{align}
H_{0,g}: \mu_{g1} = \mu_{g2} \hspace*{.25cm} \text{  vs  } \hspace*{.25cm} H_{1,g}: \mu_{g1} \ne \mu_{g2}. \nonumber
\end{align}
Per-gene mean and fitted dispersion estimates are obtained as before. In order to estimate a possible effect due to study, a full and reduced model are fitted for each gene using negative binomial generalized linear models (GLM); the full  model regresses gene expression on the experimental condition and study, while the reduced model regresses gene expression only on the study. The two models are compared using a $\chi^2$ likelihood ratio test to determine whether including the experimental condition significantly improves the model fit. Note that for the global differential analysis we use the \texttt{HTSFilter} Bioconductor package \cite{Rau2012} to filter the full set of data across studies, resulting in a vector of raw filtered per-gene $p$-values that may be corrected for multiple testing using classical procedures \cite{BH} to control the false discovery rate at a desired level $\alpha$. Additional details may be found in the \texttt{DESeq} package vignette.

\section{Results and Discussion} \label{realappli}

\subsection{Application to real data}
\label{sec:Application}

\subsubsection{Presentation of the data}

The negative binomial GLM and $p$-value combination methods were applied to a pair of real RNA-seq studies performed to compare two human melanoma cell lines \cite{Strub2011}. Each study compares gene expression in a melanoma cell line expressing the Microphtalmia Transcription Factor (MiTF) to one in which small interfering RNAs (siRNAs) were used to repress MiTF, with three biological replicates per cell line in the first (hereafter referred to as Study A) and two per cell line in the second (Study B). The raw read counts and phenotype tables for Study A are available in the Supplementary Materials of \cite{Statomique2012}, and the data from Study B from \cite{Strub2011}. 

The characteristics of the data from these two studies are summarized in Supplementary Table 2. In particular, we note that the data from Study A tend to have larger total library sizes and a smaller number of unique reads than those from Study B; in addition, Study A appears to exhibit larger overall per-gene variability than does Study B (Supplementary Figure 8). These two points indicate that in this pair of studies, a considerable amount of inter-study variability appears to be present (Supplementary Figure 9).

\subsubsection{Results}

After performing individual differential analysis for each study using the negative binomial model and exact test as described above, we obtained per-gene $p$-values for each study (Figure~1, histograms in background). As previously stated, an important underlying assumption of the $p$-value combination methods is that the $p$-values are uniformly distributed under the null hypothesis; we note that this is not the case here, especially for the second study, due to a large peak of values close to 1 resulting from the discretization of $p$-values. In order to remove the weakly expressed genes contributing to this peak in each study, we filtered the data from each study as proposed in \cite{Rau2012}, resulting in a distribution of raw $p$-values from each study that appears to satisfy the uniformity assumption under the null hypothesis (Figure~1, histograms in foreground).

\begin{figure}[h!]
\centering
\includegraphics[width=.5\textwidth]{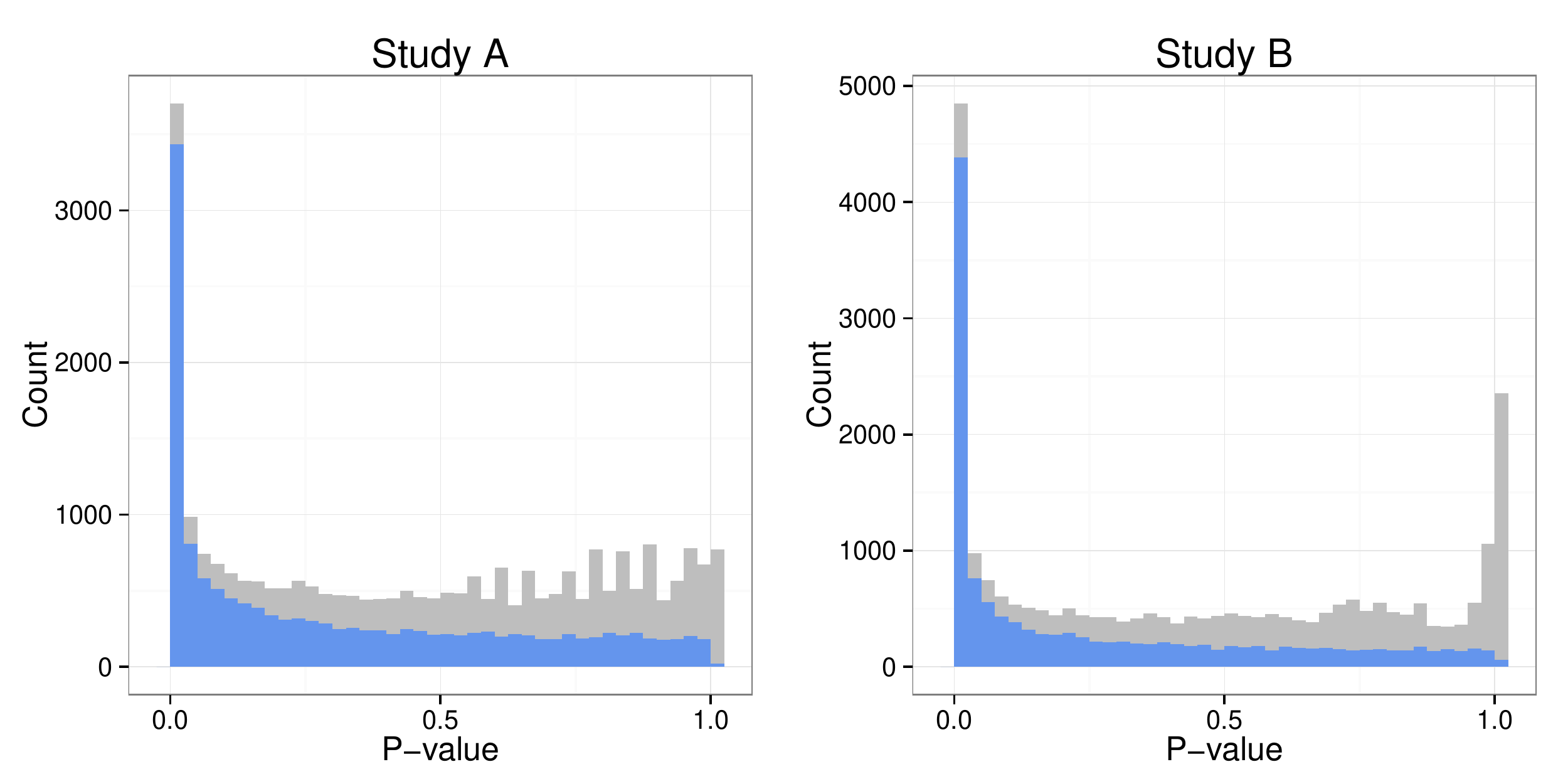}
\caption{\label{Histrawp} Histograms of raw $p$-values obtained from per-study differential analyses in the real data: unfiltered (in grey) and filtered (in blue) using the method of \cite{Rau2012}. Figure made using the \texttt{ggplot2} package \cite{Wickham2009}.}
\end{figure}

The per-study filtered $p$-values were combined using the test statistics defined in Equations~(\ref{eqn:InvNorm}) and (\ref{eqn:Fisher}), and the corresponding results were compared to those of the intersection of independent per-study analyses and the global analysis using a negative binomial GLM with a fixed study effect. We note that for the independent per-study differential analyses, a gene is declared to be differentially expressed if identified in both studies with no differential expression conflict (see Section~\ref{sec:considerations}). 

\begin{figure}
\centering
\includegraphics[width=6cm, clip=true, trim = 0 1.4in 0 1in]{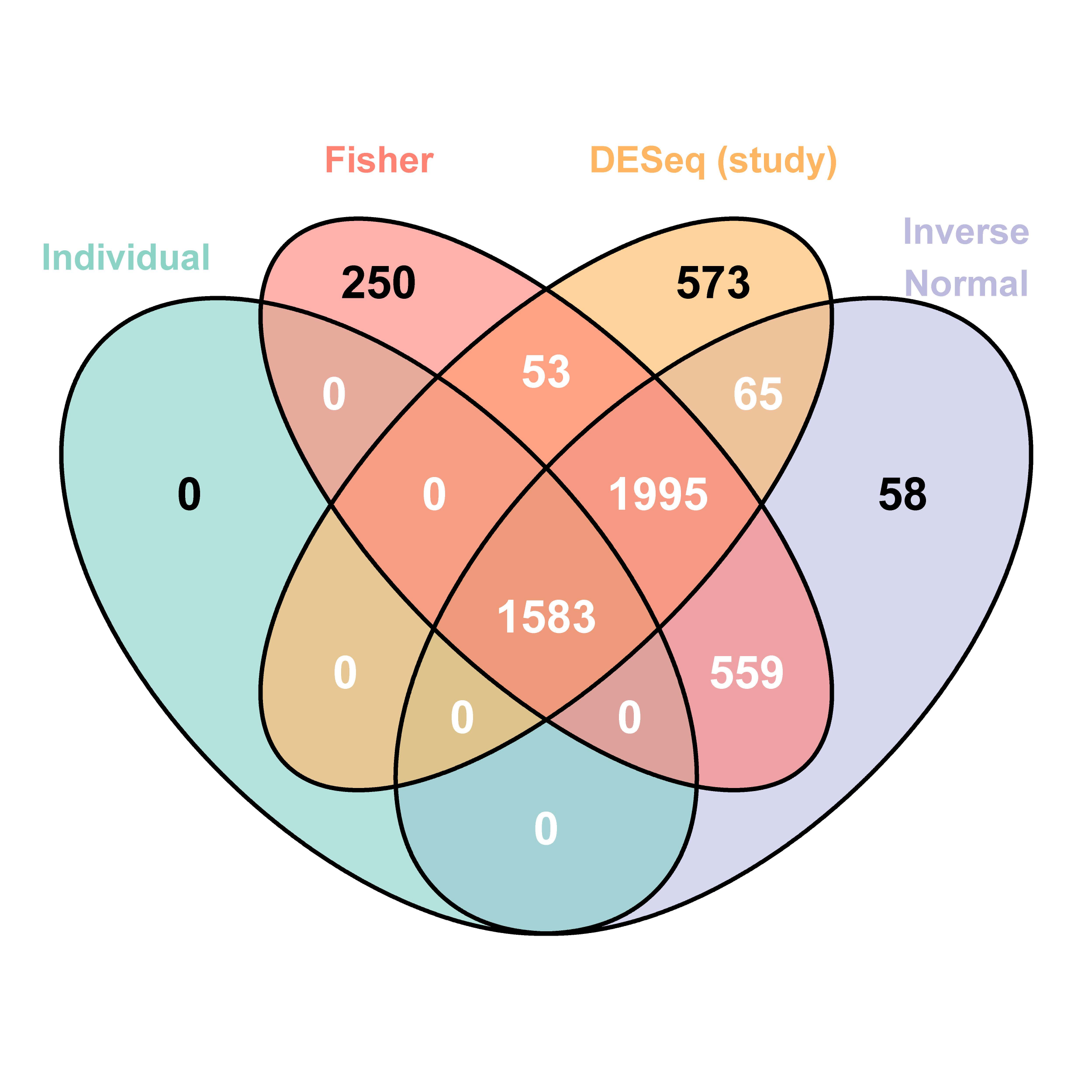}
 \caption{\label{DEresultsALL} Venn diagram presenting the results of the differential analysis for the real data for the two meta-analysis methods (Fisher and inverse normal), the global analysis (DESeq (study)), and the intersection of individual per-study analyses (Individual). Figure made using the \texttt{VennDiagrams} package \cite{Chen2012}.}
\end{figure}

The Venn diagram presented in Figure~2 compares the lists of differentially expressed genes found for all methods considered. It may immediately be noticed that the independent per-study analysis approach is very conservative, and both of the $p$-value combination approaches (Fisher and inverse normal) considerably increase the detection power.
In addition, a large number of genes are found in common among the $p$-value combination methods and the global analysis (3578 compared to only 1583 from the intersection of individual studies). In order to determine whether the genes uniquely identified by a particular method appear to be biologically pertinent, an Ingenuity Pathways Analysis (Ingenuity\textsuperscript{{\textregistered}} Systems, \href{www.ingenuity.com}{www.ingenuity.com}) was performed to identify functional annotation for the genes uniquely identified by the Fisher $p$-value combination method with respect to the global analysis, and vice versa. We note that the sets of genes uniquely identified by the Fisher method or the global analysis (Supplementary Tables 2 and 3), as well as the set of genes found in common (Supplementary Table 4), all appear to include genes of potential interest related to cancer or melanoma, which was the main focus of this set of studies. As such, for this pair of studies it appears that the union of genes identified by the two approaches may be of biological interest; to further study the effect of number of studies and inter-study variability on the performance of each method, we investigate an extensive set of simulated data in the following section.


\subsection{Simulation study}
\label{sec:sims}

Data were simulated according to a negative binomial distribution,
\begin{equation}
Y_{gcrs} \sim {\mathcal{NB}}(\mu_{gcs},\phi_{gs}) \nonumber
\end{equation}
where $\mu_{gcs}$ and $\phi_{gs}$ represent the mean and dispersion, respectively, for gene $g$, condition $c$ and study $s$, and the mean-variance relationship is defined by
\begin{equation}
\mathrm{Var}(Y_{gcrs}) = \mu_{gcs} + \frac{\mu_{gcs}^2}{\phi_{gs}}. \nonumber
\end{equation}

In order to incorporate inter-study variability, we consider the following situation for the mean parameter $\mu_{gcs}$:
\begin{equation} 
\log (\mu_{gcs}) = w_{gc} + \varepsilon_{gcs}, \text{ and } \varepsilon_{gcs} \sim {\mathcal{N}}(0,\sigma^2),\nonumber
\end{equation}
where $w_{gc}$ represents the overall mean for gene $g$ in condition $c$, $\varepsilon_{gcs}$ the variability around these means due to a study effect, and $\sigma^2$ the size of the inter-study variability. Note that as $\varepsilon_{gcs}$ affects $\mu_{gcs}$ through a log link, the value of $\exp(\varepsilon_{gcs})$ has a multiplicative effect on the mean.

\subsubsection{Parameters for simulations}

To fix realistic values for the parameters $\lbrace w_{gc}, \phi_{gs}, \sigma\rbrace$, we first performed individual per-study differential analyses by fitting a negative binomial model with the default methods and parameters of the \texttt{DESeq} package (see Section~\ref{sec:independent}) on the unfiltered human data presented in Section~\ref{sec:Application}. The per-study false discovery rate was subsequently controlled at the $\alpha = 0.05$ level \cite{BH}. For the genes identified as differentially expressed in both studies, $w_{g1}$ and $w_{g2}$ were fixed to be the values of the empirical means (after normalization for library size differences) for each condition across studies. For the remaining genes, we set $w_{g1} = w_{g2} = w_{g}$ to be the overall empirical mean (after normalization for library size differences) for gene $g$ across both conditions and studies. Using the gamma-family GLM fitted to the per-gene mean and dispersion parameter estimates for each study (Supplementary Figure 8), we fixed the dispersion parameter $\phi_{gs}$ to be equal to the fitted values
\begin{align}
\phi_{gs}^{-1} = \hat{\alpha}_{0s} + \frac{\hat{\alpha}_{1s}}{w_{g}},  \nonumber
\end{align}
where $\hat{\alpha}_{0s}$ and $\hat{\alpha}_{1s}$ are the estimated coefficients from the gamma-family GLM for study $s$, and $w_{g}$ is the overall empirical mean for gene $g$. For weakly expressed genes, it has been observed that little overdispersion is present as biological variation is dominated by shot noise (i.e., the variation inherent to a counting process); for genes with $w_g<10$, the dispersion parameter is therefore fixed to be $\phi_{gs} = 10^{10}$, which corresponds to nearly zero overdispersion (i.e., mean nearly equal to the variance).

Finally, the parameter $\sigma$ is chosen to represent a range of values for the amount of inter-study variability. The observed human data exhibit a considerable amount of inter-study variability, corresponding to a value of roughly $\sigma = 0.5$ (see Supplementary Materials, Figure 9). In the following simulations, four values are considered for the parameter $\sigma$: $\lbrace 0, 0.15, 0.3, 0.5 \rbrace$, representing zero, small, medium, and large inter-study variability, respectively. Finally, we note that for genes simulated to be non-differentially expressed, we set $\varepsilon_{g1s} = \varepsilon_{g2s} = \varepsilon_{gs} \sim \mathcal{N}(0,\sigma^2)$.

\begin{table}
\centering
\caption{Simulation settings, including the number of studies and the number of replicates per condition in each study.\label{tab:simsettings}}
\begin{tabular}{ccc}
\hline
Setting   & \# of studies & Replicates/study \\
\hline
1 & 	2 & (2,3) 		\\
2 &  	3 & (2,2,3) 	\\
3 & 	5 & (2,2,3,3,3) \\
\hline
\end{tabular}
\end{table}

The simulation settings used for the number of studies and number of replicates  per condition in each study are presented in Table~1 and were chosen to reflect the size of real RNA-seq experiments. When more than two studies were simulated, the same simulation parameters were used as for the first two, as determined from the real data. For simplicity, the same number of replicates was simulated in each condition for all studies. 

\subsubsection{Methods and criteria for comparison}
In addition to the intersection of independent per-study analyses (where genes were declared to be differentially expressed if identified in more than half of the studies with no differential conflict), the Fisher and inverse normal $p$-value combination techniques, and the global analysis with fixed study effect, we also considered a global analysis with no study effect. For each simulation setting and level of inter-study variability $\sigma$, 300 independent datasets were simulated, and the filtering method of \cite{Rau2012} was applied, either independently to each study (for the independent per-study analyses and $p$-value combination techniques) or to the full set of data (for the global analysis). 

For each method, performance was assessed using the sensitivity, false discovery rate (FDR) and area under the receiver operating characteristic (ROC) curve (AUC).
In addition, we also considered a criterion to assess the ``value added"  for the $p$-value combination methods with respect to the global analysis, and vice versa: the proportion of true positives among those uniquely identified by a given method (e.g., the Fisher approach) as compared to another (e.g., the global analysis).

\subsubsection{Results}

\begin{figure*}[t!]
\centering
\includegraphics[width=16cm]{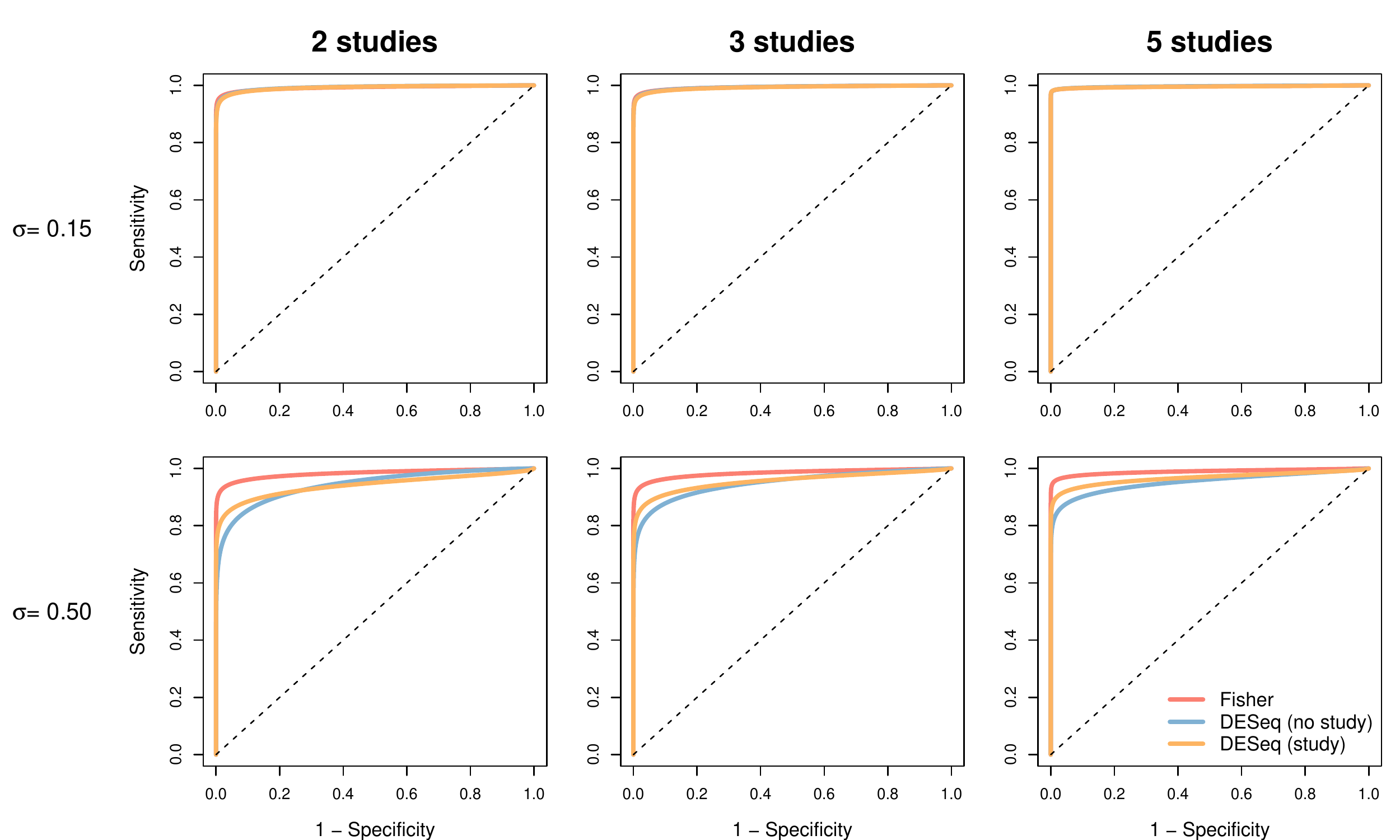}
\caption{Receiver Operating Characteristic (ROC) curves, averaged over 300 datasets. Each plot represents the results of a particular setting, with columns corresponding (from left to right) to simulations including 2 studies, 3 studies, and 5 studies, and rows corresponding (from top to bottom) to simulations with inter-study variability set to $\sigma=0.15$ and $\sigma=0.50$ (low inter-study variability to lareg inter-study variability). Within each plot:  Fisher (red lines), global analysis with no fixed study effect (DESeq (no study), blue lines), and global analysis with a fixed study effect (DESeq (study), orange lines). The dotted black line represents the diagonal.\label{fig:ROC}}
\end{figure*}

The different methods were first compared with ROC curves, presented in Figure~3 for low and high inter-study variability (results for zero and moderate inter-study variability are shown in Supplementary Figure 5). We note that for clarity, the inverse normal method is not represented on these plots as its performance was found to be equivalent to the Fisher method. It can first be noted that for no or small inter-study variability ($\sigma=0$ or $\sigma=0.15$), no practical difference may be observed among the methods. On the other hand, for moderate to large inter-study variability ($\sigma=0.3$ or $\sigma=0.5$) differences among the methods become more apparent; this pattern is observed for any number of studies. As expected, including a study effect in the global analysis improves the performance over a naive global analysis without such an effect. We note that the two proposed meta-analysis methods (inverse normal and Fisher $p$-value combination) were found to perform very similarly and were able, in the case of large inter-study variability, to outperform the global analysis in terms of AUC (Supplementary Figure 1). In particular, in the presence of large inter-study variability, the naive global analysis without a study effect unsurprisingly has the lowest AUC, and the two meta-analysis methods yield a larger AUC than the global analysis with a study effect.

\begin{figure*}
\centering
\includegraphics[width=15cm]{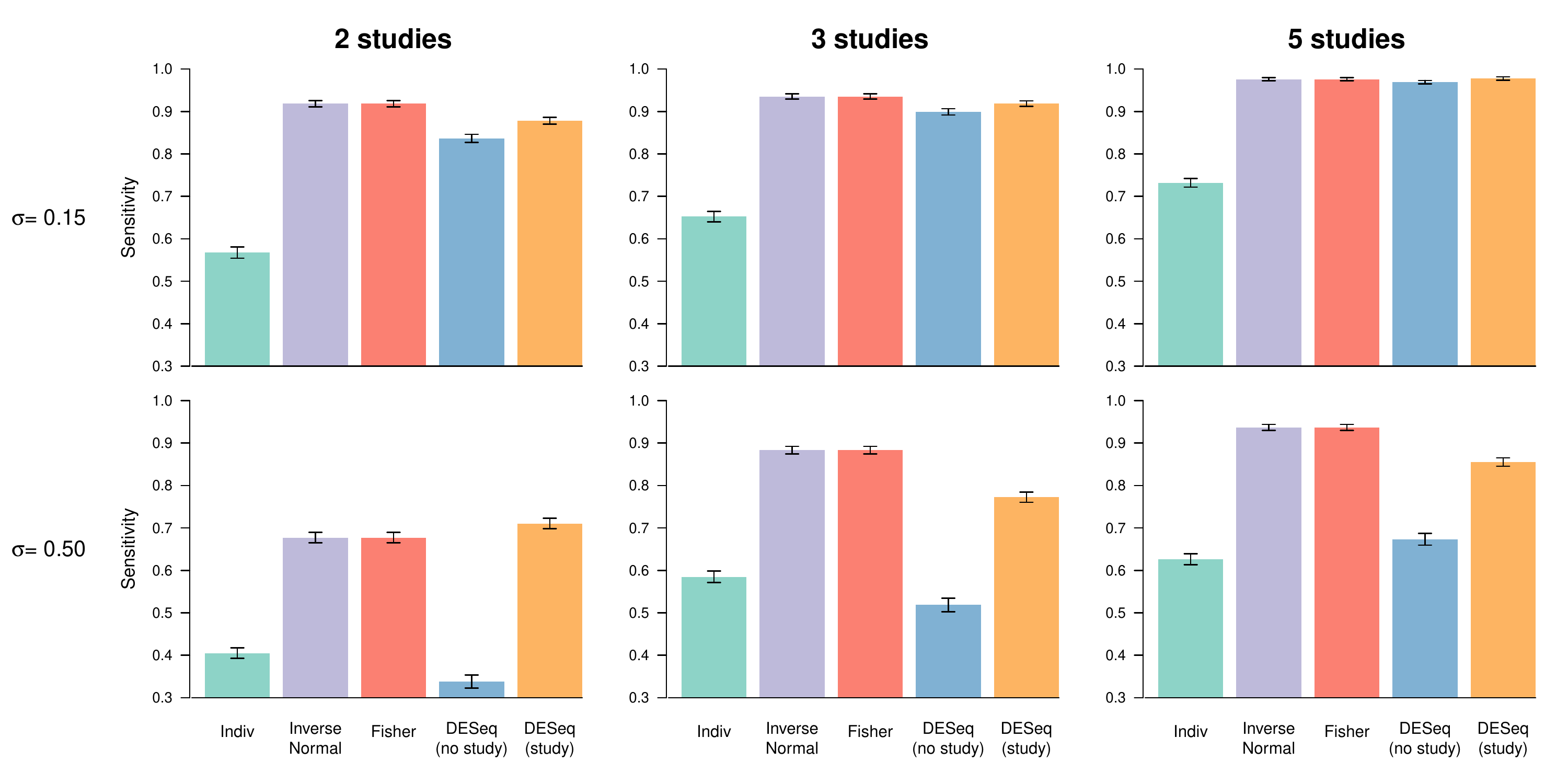}
\caption{Sensitivity, for the simulation settings corresponding to $\sigma=0.15$ and $\sigma=0.50$. Each barplot represents the results of a particular setting, with columns corresponding (from left to right) to simulations including 2 studies, 3 studies, and 5 studies, and rows corresponding (from top to bottom) to simulations with inter-study variability set to $\sigma=0.15$ and $\sigma=0.50$ (no inter-study variability to moderate inter-study variability). Within each barplot, from left to right: Individual per-study analyses (green bars), Inverse Normal (purple bars), Fisher (red bars), DESeq with no study effect (blue bars), and DESeq with a fixed study effect (orange bars).  \label{fig:sensitivity}}
\includegraphics[width=10cm]{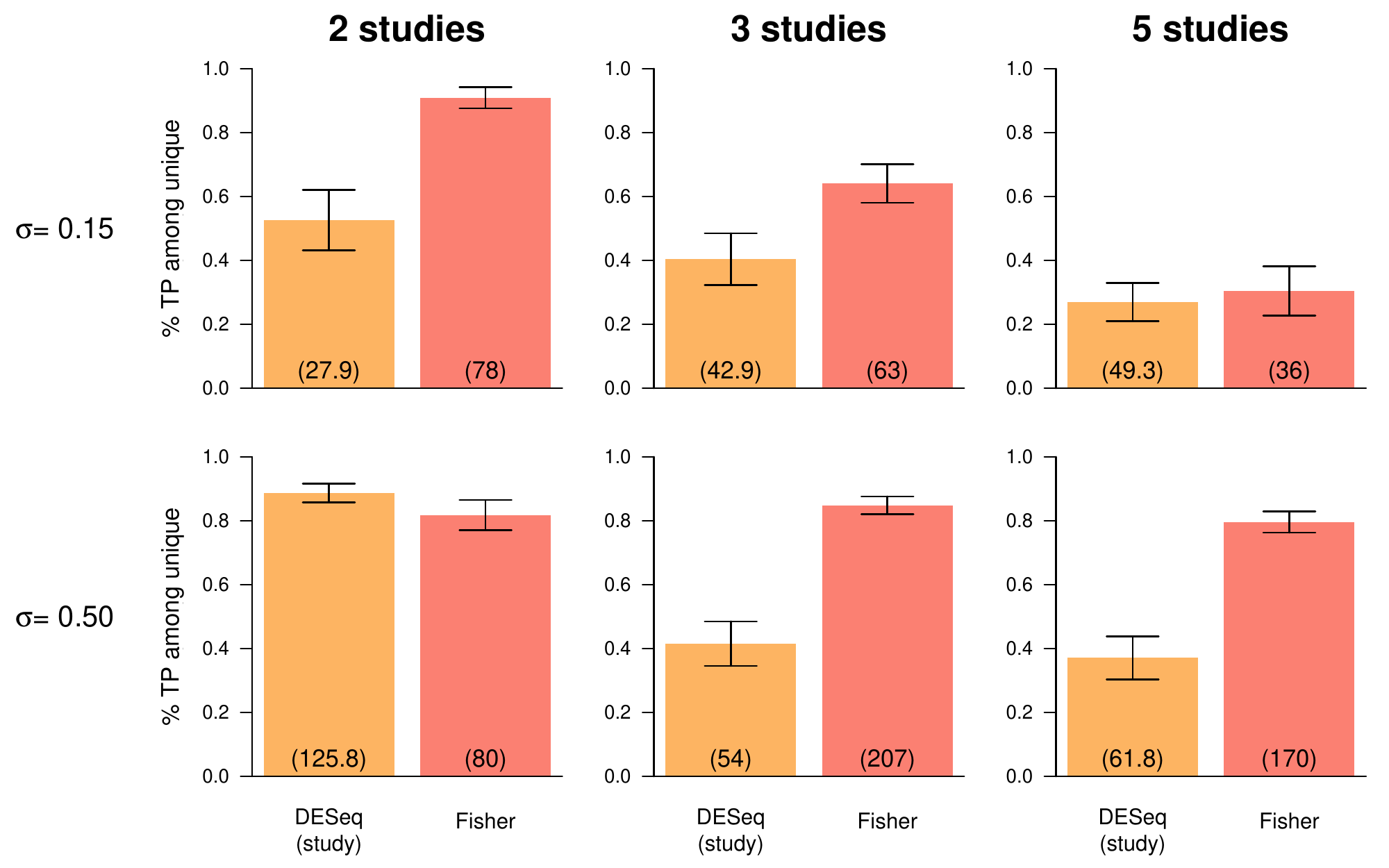}
\caption{Proportion of true positives among unique discoveries for DESeq with a fixed study effect (orange bars) and Fisher (red bars) for the simulation settings not demonstrated in the main paper. Each barplot represents the results of a particular setting, with columns corresponding (from left to right) to simulations including 2 studies, 3 studies, and 5 studies, and rows corresponding (from top to bottom) to simulations with inter-study variability set to $\sigma=0.15$ and $\sigma=0.50$ (no inter-study variability to moderate inter-study variability). Error bars represent one standard deviation, and numbers in parentheses represent the mean total number of unique discoveries for DESeq with study effect as compared to Fisher and vice versa, respectively. \label{fig:TPdiscoveries}}
\end{figure*}

Considering the sensitivity (Figure~4 and Supplementary Figure 6), the meta-analyses appear to lead to similar, and in some settings considerably higher, detection power compared to the other methods. We note that in all settings, using the intersection of independent analyses leads to much lower sensitivity, even for low or zero inter-study variability. As for the AUC, the sensitivity was found to be considerably improved for the global analysis when including a study effect in the GLM model, particularly for medium to large inter-study variability. The two meta-analysis methods were found to lead to significant improvements in sensitivity as compared to the global analysis in the presence of moderate to large inter-study variability when three or more studies were considered.  However, for the setting that most resembles our real data analysis (2 studies, $\sigma=0.50$), the global analysis with study effect and meta-analyses appear to have similar detection power. Finally, we also note that for all methods the FDR was well controlled below 5$\%$ (Supplementary Figure 2). 

Based on these criteria, the two proposed meta-analysis methods (inverse normal and Fisher) seem to perform very similarly. In order to more thoroughly investigate the differences between $p$-value combination methods and the global analysis including a study effect, we calculated the proportion of true positives uniquely detected by the Fisher method as compared to the global analysis with study effect, and vice versa (Figure~5 and Supplementary Figure 7). In the setting closest to the real data analysis presented above (two studies and large inter-study variability), the proportion of true positives found uniquely by either the Fisher approach or the global analysis with fixed study effect are very large (around 80$\%$ for both methods). This seems to suggest that the additional genes uniquely found either by the global analysis or Fisher $p$-value combination method in the real data application may indeed be of great biological interest. For more than two studies, however, as the inter-study variability increases the proportion of truly differentially expressed genes uniquely found by the Fisher method increases compared to the global analysis. For example, for three studies with large inter-study variability ($\sigma=0.5$), the proportion of truly DE genes uniquely found with the Fisher method was equal to more than 80$\%$, whereas it was only around 40$\%$ for the global analysis with a study effect.


\section{Conclusions} \label{discussion}

The aim of this paper was to present and compare different strategies for the differential meta-analysis of RNA-seq data arising from multiple, related studies. As expected, naive analyses such as the overlap of lists of differentially expressed genes found by individual studies or a global analysis not accounting for a study effect perform very poorly. On the other hand, the two proposed meta-analysis methods seem to have very similar performances. For low inter-study variability, the results are very close to those of a global GLM analysis including a study effect. When the inter-study variability increases, however, the gains in performance in terms of AUC, sensitivity, and proportion of true positives among uniquely identified genes for the meta-analysis techniques are significant as compared to the global analysis, particularly for the analysis of data from more than two studies. We note that both of the proposed $p$-value combination methods are implemented in an R package called  \href{https://r-forge.r-project.org/R/?group_id=1504}{\texttt{metaRNASeq}}, available on the R-Forge. 

Our focus in this work is on differential analyses between two experimental conditions, but can readily be extended to multi-group comparisons. However, as previously noted, the methods presented here are intended for the analysis of data in which all experimental conditions under consideration are included in every study, thus avoiding problems due to the confounding of condition and study effects. As with all meta-analyses, the $p$-value combination techniques presented here must overcome differences in experimental objectives, design, and populations of interest, as well as differences in sequencing technology, library preparation, and laboratory-specific effects.

The differential meta-analyses presented here concern expression studies based on RNA-seq data. However, other genomic data are generated by high-throughput sequencing techniques, including chromatin immunoprecipitation sequencing (CHIP-seq) and DNA methylation sequencing (methyl-seq), and the proposed techniques could potentially be extended to these other kinds of data. However, in order to be biologically relevant, the $p$-value combination methods rely on the fact that the same test statistics, or in the case of RNA-seq data conditioned tests, are used to obtain $p$-values for each study. An important challenge for the future will be to propose methods able to jointly analyze related heterogeneous data, such as microarray and RNA-seq data, or other kinds of genomic data. This is not straightforward in a meta-analysis framework and remains an open research question.

\bigskip
\section*{Authors' contributions}
AR participated in the design of the study, performed simulations and data analyses, and helped draft the manuscript.
GM designed the study, wrote the associated R package, and helped draft the manuscript.
FJ conceived of the study, participated in its design, and drafted the manuscript.
All authors read and approved the final manuscript.

\section*{Acknowledgements}
{\small
We thank Thomas Strub, Irwin Davidson, C\'eline Keime and the IGBMC sequencing platform for providing the RNA-seq data.}

\newpage
\bibliographystyle{abbrv}
{\small \bibliography{MetaRNASeq}}

\end{document}